\documentclass[fleqn,usenatbib]{mnras}
\usepackage{newtxtext,newtxmath}

\usepackage{epsfig}
\usepackage{amsmath,color,bm}
\usepackage{booktabs}
\usepackage{subfig}

\makeatletter
\renewcommand{\p@subfigure}{}
\makeatother
\usepackage{nameref}
\usepackage{graphicx}
\usepackage{graphics}
\usepackage[space]{grffile}
\usepackage{url}
\usepackage[utf8]{inputenc}
\usepackage{fancyref}
\usepackage{xcolor}
\usepackage[normalem]{ulem}
\usepackage{eqnarray,amsmath}

\title[Improved Early-Warning with Higher Modes]{Improved early warning of compact binary mergers using higher modes of gravitational radiation: A population study }

\author[M. K. Singh et al.]{Mukesh Kumar Singh,$^1$
  Shasvath J. Kapadia,$^1$
  Md Arif Shaikh,$^1$
  Deep Chatterjee$^{2,3,4}$ \newauthor
  and Parameswaran Ajith$^{1,5}$\\
  $^1$International Centre for Theoretical Sciences, Tata Institute of Fundamental Research, Bangalore 560089, India\\
  $^2$Center for AstroPhysical Surveys, National Center for Supercomputing Applications, Urbana, IL, 61801, USA\\
  $^3$Department of Astronomy, University of Illinois at Urbana-Champaign, 1002 W. Green St., IL 61801, USA\\
  $^4$Illinois Center for Advanced Studies of the Universe, Department of Physics, University of Illinois at Urbana-Champaign, Urbana, IL 61801, USA\\
  $^5$Canadian Institute for Advanced Research, CIFAR Azrieli Global Scholar, MaRS Centre, West Tower, 661 University Ave, Toronto, ON M5G 1M1, Canada}

\begin{document}
\label{firstpage}
\pagerange{\pageref{firstpage}--\pageref{lastpage}}
\maketitle

\begin{abstract}
A gravitational-wave (GW) early-warning of a compact-binary coalescence event, with a sufficiently tight localisation skymap, would allow telescopes to point in the direction of the potential electromagnetic counterpart before its onset. Use of higher-modes of gravitational radiation, in addition to the dominant mode typically used in templated real-time searches, was recently shown to produce significant improvements in early-warning times and skyarea localisations for a range of asymmetric-mass binaries. We perform a large-scale study to assess the benefits of this method for a population of compact binary merger observations. In particular, we inject 100,000 such signals in Gaussian noise, with component masses $m_1 \in \left[1, 60 \right] M_{\odot}$ and $m_2 \in \left [1, 3 \right] M_{\odot}$. We consider three scenarios involving ground-based detectors: the fifth (O5) observing run of the Advanced LIGO-Virgo-KAGRA network, its projected Voyager upgrade, as well as a proposed third generation (3G) network. We find that for fixed early warning times of $20-60$ seconds, the inclusion of the higher modes can provide localisation improvements of a factor of $\gtrsim 2$ for up to $\sim 60\%$ ($70 \%$) of the neutron star-black hole systems in the O5 (Voyager) scenario. Considering only those neutron star-black hole systems which can produce potential electromagnetic counterparts, such improvements in the localisation can be expected for $\sim 5-35\%$ $(20-50\%)$ binaries in O5 (Voyager). For the 3G scenario, a significant fraction of the events have time gains of a minute to several minutes, assuming fiducial target localisation areas of 100 to 1000 sq. deg.        
\end{abstract}

\begin{keywords}
Gravitational waves
\end{keywords}

\section{Introduction}\label{sec:introduction}
	Multimessenger detections of compact binary mergers have the potential to answer longstanding astrophysical questions, and shed light on complex phenomena that require observational data to guide and constrain the corresponding theoretical models. This potential was spectacularly demonstrated with GW170817 \citep{GW170817-DETECTION}--- the binary neutron star (BNS) merger that was first detected by the LIGO-Virgo network \citep{advligo, advvirgo} and then followed up extensively by electromagnetic telescopes worldwide \citep{GW170817-MMA, O2GWMMA}. Apart from being the very first observation of a BNS merger with an electromagnetic counterpart, GW170817 also revealed BNS mergers to be engines of short gamma-ray bursts \citep[see][for a review]{NakarGRB} and kilonovae \citep{MetzgerKN}, and sites where heavy elements get synthesized \citep{GW170817HeavyElements}. It further allowed a hitherto unprecedented probe of the equation of state (EOS) of ultra-dense nuclear matter \citep{GW170817-EOS, GW170817-SOURCE-PROPERTIES}, a remarkably stringent test of the equality between the speed of gravitational waves and the speed of light \citep{GW170817-TGR, Liu2020}, as well as a distance-ladder-independent measurement of the Hubble constant to complement existing measurements \citep{GW170817-HUBBLE}. An early-warning of such electromagnetically-bright (EM-Bright) BNS and neutron star-black hole (NSBH) events would further shed light on complex physics surrounding the merger, including precursors \citep{InspiralCounterpart} and short-lived intermediate merger products \citep{HotokezakaHNS}. 

In the recently-completed third observing run (O3) of Advanced LIGO and Virgo, estimates of the significance, source-classification \citep{KapadiaPastro}, and source-properties \citep{ChatterjeeEMB} of candidate events were reported for the first time in low-latency \citep{user_guide}. Among the exceptional events detected during this observing run is an unusually heavy [by galactic BNS standards \citep{Farrow2019}]\footnote{Note that the possibility of this event being an unusually light binary black hole merger can not be ruled out.} putative BNS merger \citep{GW190425-detection}, as well as a compact binary merger whose estimated component masses are not inconsistent with a NSBH system \citep{GW190814-detection}. However, no confirmed detection of electromagnetic counterparts were announced for any of the O3 events \footnote{An intriguing candidate electromagnetic counterpart to GW190521 event in O3 was reported by the Zwicky Transient Facility \citep{Graham2019}. It has been speculated that the candidate may be confirmed or rejected based on future observations  \citep{ZTF-BBH}}. 

Nevertheless, galvanized by the immense science gains of GW170817's multimessenger detection, and the plans of a significantly improved sensitivity of the LIGO-Virgo network \citep{advligo, advvirgo} enhanced by the KAGRA detector \citep{kagra} for the next observing run \citep{observer_summary}, efforts towards setting up a GW early-warning system are currently under way \citep{Sachdev2020, NitzEW, SpiirEW}. These efforts are primarily targeted towards BNS systems traditionally expected to produce counterparts. These efforts essentially rely on the relatively long duration of the BNS inspiral (to be precise, the dominant mode of the corresponding GW signal) in the frequency band of the detectors, thus allowing real-time templated searches to accumulate sufficent signal-to-noise (SNR) tens of seconds to a minute before the merger \citep{CannonEW}.

Early-warning of heavier compact binary systems, such as NSBHs or BBHs, are unlikely to benefit from methods relying solely on the time spent by the dominant mode of the gravitational radiation in the frequency band ($\sim 10-2000$ Hz) of ground based detectors. However, this impediment can be circumvented if the subdominant modes are sufficiently excited, as is the case for asymmetric-mass systems and a range of orbital inclination angles. Since higher-modes vibrate at higher harmonics of the orbital frequency than the dominant mode, they will enter the detectors' frequency band well before the dominant mode --- a fact that can be used to enhance early-warning targeted at heavier asymmetric mass systems.

In our previous work \citep{Kapadia2020}, we showed that inclusion of some of the subdominant modes ($\ell = m = 3, \ell = m = 4$), in addition to the dominant mode ($\ell = m = 2$) conventionally used in low-latency GW searches \citep{gstlal, mbta, pycbc, spiir}, can improve the early-warning times and reduce the localisation skyareas, for a range of asymmetric-mass binaries and orbital inclinations. We considered three observing scenarios: O5, with five detectors including LIGO-India \citep{aplus_sensitivity, observer_summary,LIGO-India}; Voyager, with the same five detectors, but with the three LIGO detectors upgraded to Voyager sensitivity \citep{Voyager_PSD, Adhikari_voy}; and a 3G scenario consisting of two Cosmic Explorers and one Einstein Telescope \citep{CE, ET, CE_PSD, ET_PSD}. For the O5 (Voyager) scenario, NSBH mergers located at 40 Mpc, including those that have the potential to produce EM- counterparts, can be localised to hundreds of square degrees in the sky as early as 45 seconds before merger. This corresponds to a reduction factor of up to $3-4(5-6)$. For the 3G scenario, potentially EM-Bright mergers located at 100 Mpc could have early-warning time gains of up to $\mathcal{O}$(minutes) corresponding to a localisation area of 100 sq. deg. The exact fraction of detectable NSBH events that benefit from inclusion of higher modes depends on the population distributions, which we study in this work.

This sequel to our previous work attempts to assess the benefits of the inclusion of higher modes in a statistical sense. In particular, our goal is to determine what fraction of events of a fiducial population is likely to benefit (in terms of early-warning gains) from the inclusion of higher modes. We inject a population of 100,000 compact binaries, with masses $m_1 \in \left [1, 60 \right ] M_{\odot}$ and $m_2 \in \left [1, 3 \right ] M_{\odot}$ log-uniformly distributed, and distances distributed uniformly in co-moving volume. We consider three limiting distances: $40$ Mpc, $100$ Mpc and $200$ Mpc, motivated by typical distances at which EM telescopes can view the EM-counterparts of compact binary mergers with sufficiently short exposure times. We also consider the three observing scenarios mentioned above. Additionally, we focus on sub-populations of the simulated compact binaries, such as NSBH systems, as well as sub-populations of NSBH systems that could produce EM counterparts under different assumptions of the equation of state of the neutron star and spin distribution of the black hole \citep{FoucartEMB1, FoucartEMB2}. We find that for O5 (Voyager), up to $\sim 60 \% (\sim 70 \%)$ of the NSBH population provide skyarea reduction factors of $\gtrsim 2$ with the inclusion of higher modes for an early warning time of $60$ seconds, whereas the localisation area itself depends on the distance to the source. Depending on the choice of equation of state of neutron star and spin distribution of the black hole, the fraction of EM-Bright events with reduction factors $\gtrsim 2$ can vary by $10 \% - 20 \%$. Early-warning time gains for fixed localisation areas ($100-1000$ sq. deg.) are explored for the 3G scenerio, with a large fraction of the total population producing early warning time gains of a minute or more. Additionally, early-warning time gains for fixed SNRs are also estimated for all observing scenarios.


The rest of the paper is organized as follows. Section~\ref{sec:method} motivates the need for a large-scale study, as well as summarizes the method. Section~\ref{sec:results} summarizes the results and Section~\ref{sec:conclusion} gives an overview of the study conducted as well as broad conclusions drawn from the study which advocate the use of higher modes in real-time GW searches to enhance their early-warning abilities.

\section{Motivation and Method}\label{sec:method}
\subsection{Calculating the expected localisation skyarea} 
Working with the assumption that the GW detector noise is stationary and Gaussian, determining a trigger to be significant enough for follow-up comes down to setting some pre-defined threshold on the matched filter SNR. This is because the statistical properties of zero-mean, stationary Gaussian noise are entirely determined by the power spectral density (PSD), $S(f)$, and the optimal filter to search for known signals buried in it is the matched filter. The localisation area pertaining to a confidence interval can be estimated from the separation of the detectors, their individual effective bandwidths, and the SNRs. We briefly discuss below the method \footnote{Note that this is a Fisher matrix based approach which provides a lower bound on statistical errors.} to evaluate the localisation skyarea, which is based entirely on \citet{Fairhurst1, Fairhurst2}; the reader may refer to these for additional details.

It is convenient to express the SNR and the bandwidth of a detector in terms of the frequency moments:
\begin{eqnarray}
\overline{f^n} = 4\int_0^{\infty}f^n\frac{\mid{h(f)}\mid^2}{S(f)}df
\end{eqnarray}
where $h(f)$ is the Fourier transform of the GW waveform.
The SNR ($\rho$) is computed from the zeroth ($n = 0$) moment (the network SNR is the quadrature sum of the individual detector SNRs), and the effective bandwidth ($\sigma_f$) is computed from the first ($n = 1$) and second ($n = 2$) moments:
\begin{eqnarray}
\rho^2 &=& \overline{f^0} \\
\sigma_f^2 &=& \frac{1}{\rho^2}\overline{f^2} - \left(\frac{1}{\rho^2}\overline{f}\right)^2
\end{eqnarray}
The uncertainty ($\sigma_t$) in the time of arrival of the GW at a detector can be computed from the SNR and the bandwidth as:
\begin{eqnarray}
\sigma_t = \frac{1}{2\pi\rho\sigma_f}
\label{eq:timing_uncertainty}
\end{eqnarray}
%
Working in an earth centered coordinate system, we define the source position by $\textbf{\emph{R}}$ and the time at which the signal passes through the center of the earth as $T_{o}$. $\textbf{\emph{R}}$ is a unit vector since we are interested in sky-location of the source and not in distance. If we denote the location of the detector $i$ with respect to the center of the earth by $\textbf{\emph{d}}_{i}$ (in units of time by dividing by the speed of light), the time at which the signal passes through the detector $i$ is given by:
\begin{eqnarray}
T_i = T_o - \textbf{\emph{R}}\cdot \textbf{\emph{d}}_i
\label{eq:time_detector_i}
\end{eqnarray}
Then, the probability of the measured value of arrival times $t_i$ in different detectors, given the true arrival times $T_i$, is given by:
\begin{eqnarray}
p(t_i|T_i) = \prod_i \frac{1}{\sqrt{2 \pi} \sigma_i} \exp \left[\frac{-(t_i - T_i)^2}{2 \sigma_i^2} \right]
\label{eq:dist_t_i}
\end{eqnarray}
where $\sigma_i$ is the timing uncertainty defined in Eq. (\ref{eq:timing_uncertainty}). We have assumed that timing errors are Gaussian distributed. Eq. (\ref{eq:dist_t_i}) is nothing but the likelihood for the measured value of arrival times $t_i$ given their true values. Using Bayes' theorem, we obtain the posterior for the true arrival times $T_i$ assuming a prior $p(T_i)$.

Since our final goal is to obtain a posterior on the sky-location of the event, we need to express the posterior $p(T_i \mid t_i)$ in terms of $\textbf{\emph{R}}$. To that end, we write down the relation between the measured sky position $\textbf{\emph{r}}$ and measured arrival time $t_i$ in analogy with Eq.(\ref{eq:time_detector_i}).
We choose prior distributions to be uniform over the sphere (for $\textbf{\emph{R}}$) and uniform in time (for $T_o$). The posterior distribution for sky-position is then given by:
 \begin{eqnarray}
  p(\textbf{\emph{R}}|\textbf{\emph{r}}) \propto p(\textbf{\emph{R}}) \exp \left[- \frac{1}{2} (\textbf{\emph{r}} - \textbf{\emph{R}})^\mathrm{T} \boldsymbol{\mathsf{M}} (\textbf{\emph{r}} - \textbf{\emph{R}}) \right]
 \label{eq:post_R}
 \end{eqnarray}
where $\mathrm{T}$ is the matrix transpose. The Fisher matrix $\boldsymbol{\mathsf{M}}$, describing the localization uncertainty, is given by:
 \begin{eqnarray}
  \boldsymbol{\mathsf{M}} = \frac{1}{\sum_k \sigma_k^{-2}} \sum_{i,j} \frac{\boldsymbol{\mathsf{D}}_{ij} \boldsymbol{\mathsf{D}}_{ij}^T}{2 \sigma_i^2 \sigma_j^2}
 \end{eqnarray}
and $\boldsymbol{\mathsf{D}}_{ij} = \textbf{\emph{d}}_i - \textbf{\emph{d}}_j$ represents the separation between $i^{th}$ and $j^{th}$ detectors. The localisation area at a given confidence, pertaining to the posterior on ${\bf R}$, can then be readily computed from $\boldsymbol{\mathsf{M}}$. The pre-factor in matrix $\boldsymbol{\mathsf{M}}$, is nothing but the square sum of timing uncertainties in all detectors in the network, which arises while marginalising over geocentric time $T_0$.

\subsection{Observing Scenarios}

\begin{figure}
  \begin{center}
    \includegraphics[width=1\columnwidth]{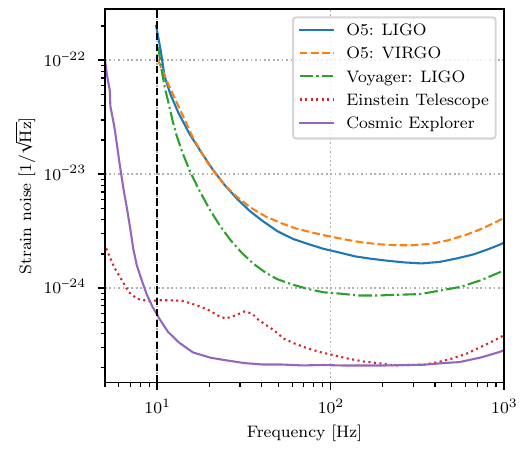} 
  \end{center}
  \caption{Projected noise amplitude spectral densities for the three LIGO detectors (including LIGO India) and the Virgo detector for the O5 and Voyager scenarios, as well as for the two Cosmic Explorers and the Einstein Telescope. The lower limit on the detector bandwidths for O5 and Voyager is set to $10$ Hz. For 3G, this limit is set to $5$ Hz. (See \citep{aplus_sensitivity, observer_summary, CE, ET, Voyager_PSD, CE_PSD, ET_PSD})
  }
  \label{fig:psd}
\end{figure}

We consider three observing scenerios involving networks of ground based interferometric detectors. {\it O5}: The fifth observing run involves a network of five detectors, consisting of the three LIGO detectors (LIGO Hanford, LIGO Livingston and LIGO-India) with an $\mathrm{A}+$ sensitivity, which corresponds to a BNS range of $330$ Mpc; the Virgo detector with a BNS range of $260$ Mpc; and a KAGRA detector assumed to have a sensitivity similar to the Virgo sensitivity \citep{aplus_sensitivity, observer_summary}. {\it Voyager}: In the Voyager scenario, we assume that all the LIGO detectors, including LIGO-India, are upgraded to Voyager sensitivity, which corresponds to a BNS comoving range of $1100$ Mpc, and the Virgo and KAGRA detectors are upgraded to $\mathrm{A}+$ sensitivity \citep{Adhikari_voy, Voyager_PSD, instrument_science}. {\it 3G}: The third generation network consists of two Cosmic Explorers (BNS comoving range of $4200$ Mpc) with geographical coordinates identical to those of LIGO Hanford and LIGO Livingston, and an Einstein telescope (in its L-shaped configuration, and BNS range similar to Cosmic Explorer) with coordinates equal to that of Virgo \citep{instrument_science, CE_PSD, ET_PSD}. The projected noise amplitude spectral densities of these detectors, pertaining to the three observing scenerios, are plotted in Fig.~\ref{fig:psd}. The lower limit on the detector bandwidths for O5 and Voyager is set to $10$ Hz. For 3G, this limit is set to $5$ Hz. 

\subsection{Higher modes of gravitational radiation}
Gravitational waves from the coalescence of compact binaries, can be written as a complex combination $h(t) := h_{+}(t) - ih_{\times}(t)$ of two polarizations $h_{+} (t)$ and $h_{\times}$($t$). This can then be expanded in the basis of spin $-2$ weighted spherical harmonics $Y^{-2}_{\ell m}(\iota, \varphi_{o})$ \citep{NewmanPenrose}:
%
\begin{eqnarray}
h(t; \iota, \varphi_{o}) = \frac{1}{d_{L}} \sum_{\ell = 2}^{\infty}\sum_{m = -\ell}^{\ell} h_{\ell m}(t, \blambda) Y^{-2}_{\ell m}(\iota, \varphi_{o})
\end{eqnarray}
Here, $d_L$ is the luminosity distance which scales the amplitude of the GW, and $h_{\ell m}$ represent various multipoles of the gravitational radiation. These are purely dependent on the intrinsic parameters ($\blambda$) of the source (such as masses and spin angular momenta of the compact objects) and time $t$, whereas the angular dependence of the waveform is inherited by the spin$-2$ weighted spherical harmonics $Y^{-2}_{\ell m}(\iota, \varphi_{o})$. Here, the polar angle $\iota$ (the azimuthal angle $\varphi_{o}$ can be set to zero by an appropriate choice of source-centered axes) defines the orientation of the angular momentum of the binary with respect to the line of sight of the observer.

The dominant contribution to the amplitude of the GW comes from the quadrupole mode, i.e. $(\ell=2,m =\pm 2)$. The next two subdominant multipoles are $(\ell=3,m =\pm 3)$ and $(\ell=4,m =\pm 4)$. The contribution of subdominant modes relative to the quadrupole mode depends on the symmetries of the system. For example, GWs from high mass ratio and high inclination angle binaries have relatively strong imprints from the subdominant modes (see, e.g. \cite{Varma2014}). 

Online real-time templated searches for compact-binary-coalescences \citep{gstlal, mbta, pycbc, spiir} employ banks of waveform-templates that only use the dominant multipole \citep{taylorf2, EOB_model, Phenom_model}. However, some of the recent LIGO-Virgo events, such as GW190412 and GW190814, were found to produce confident detections of subdominant modes in offline searches \citep{GW190412-detection, GW190814-detection, Roy2020} \footnote{Note that, for non-pressing binaries, where the orbital motion is restricted to a plane, the negative $m$ modes are related with the positive $m$ modes corresponding to the relation, $h_{\ell,-m}=(-1)^{\ell}h^{*}_{\ell m}$. In this paper, we do not consider precession, so the above relation is applicable.}.

The instantaneous frequency at which each multipole oscillates is proportional to the orbital frequency of the binary (assuming a non-precessing orbit):
\begin{eqnarray}
F_{\ell m}(t) \simeq mF_\mathrm{orb}(t)
\label{eqn:Fourier_freq_relation}
\end{eqnarray}
with $-\ell \leq m \leq \ell$, and $\ell \geq 2$ for GWs in General Relativity. Thus, higher modes (with $m > 2$) enter the frequency band of the detectors at fixed frequency intervals before the dominant mode; this is true regardless of the parameters of the binary. The time to coalescence of a binary system can be approximated, when it has reached an orbital frequency of $F_{\mathrm{orb}}$, by the expression \citep{Sathyaprakash1994}:
\begin{eqnarray}
\tau \simeq \frac{5}{256} \mathcal{M}^{-5/3}(2 \pi F_\mathrm{orb})^{-8/3} \propto (F_{\ell m}/m)^{-8/3}
\end{eqnarray}
where $\mathcal{M}:= (m_1 m_2)^{3/5}/(m_1 + m_2)^{1/5}$ is the chirp mass of the binary and $F_{\mathrm{orb}}$ is replaced in terms of $F_{\ell m}$ using Eq. (\ref{eqn:Fourier_freq_relation}). It is clear that in-band duration of a multipole $h_{\ell m}$ gains a factor of $(m/2)^{8/3}$ as compared to the quadrupole mode. Thus, this increase is $\sim 3$ and $\sim 6$ fold for the $\ell = m = 3$ mode and $\ell = m = 4$ mode respectively.

\begin{figure}
  \begin{center}
    \includegraphics[width=1\columnwidth]{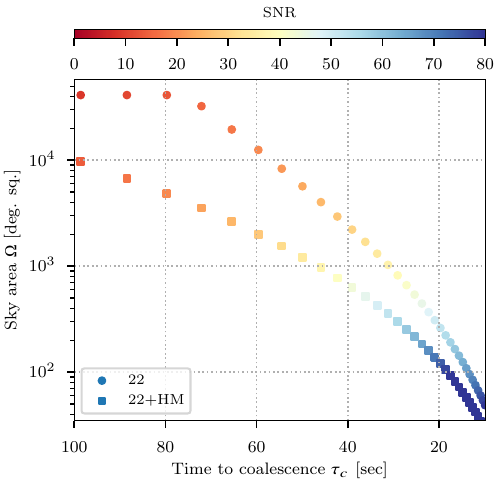} 
  \end{center}
  \caption{The evolution of the localisation skyarea and SNR as a function of the time to coalescence for a non-spinning binary merger with $m_1 = 15 M_{\odot}, m_2 = 1.5 M_{\odot}$, $d_L = 40$ Mpc, $\iota = 60$ deg. and sky location at its optimal value (that minimises the skyarea), assuming the O5 scenario. The localisation skyarea is consistently smaller, and the SNR consistently larger, with the inclusion of higher modes in addition to the dominant mode (shown by square markers), than when only the dominant mode is used in the template waveform (shown by circles). 
  }
  \label{fig:skyarea-tc}
\end{figure}

The increased in-band duration due to inclusion of these two subdominant modes, in addition to the dominant mode, can be used to acquire a given localisation skyarea and SNR at earlier times in the inspiral as compared to when only the dominant mode is used in the template waveform. This is illustrated in Fig.~\ref{fig:skyarea-tc} for a non-spinning binary merger with $m_1 = 15 M_{\odot}, m_2 = 1.5 M_{\odot}$, $d_L = 40$ Mpc and $\iota = 60$ deg in the O5 scenario described in the previous subsection.

\subsection{Relation to Previous Work}
In \citet{Kapadia2020}, we showed that the source localisation of a GW compact binary coalescence event, for fixed early-warning times, can be significantly improved with the inclusion of the $\ell=m=3$ and $\ell=m=4$ modes, in addition to the $\ell=m=2$ mode generally used in real-time searches. However, this improvement is not uniform across the space of component masses. There are two competing effects at play. On the one hand, the more asymmetric the system, the louder the subdominant modes become, effectively increasing the duration of the detectable higher-mode signal in the band of the detectors. On the other hand, the more asymmetric the system, the heavier it tends to become, which decreases the duration of the inspiral in-band. There is therefore an optimal region in the component-space where this method provides its most significant advantages; this region depends on the sensitivity of the detector network, as well as the luminosity distance to the sources. 

Given that very little is known about asymmetric-mass compact binary mergers, a range of which could produce EM counterparts \citep{FoucartEMB1, FoucartEMB2}, it may be argued that the inclusion of higher modes in real-time searches could be extremely useful, even if the advantages aren't uniformly significant across the component mass space. Nevertheless, it would be of interest to perform a quantitative study to evaluate what fraction of a fiducial population of compact binaries, which include BNS and NSBH systems, are expected to provide significant early-warning gains with the inclusion of higher modes.

\section{Results}\label{sec:results}
	We simulate 100,000 synthetic GWs \footnote{We use the \textsc{IMRPhenomHM} \citep{imrphenomhm} approximant as implemented in the \textsc{lalsuite} software package \citep{lalsuite}} for a population of compact binaries and inject them in stationary Gaussian noise. The component masses, distributed log-uniformly, span $m_1 \in \left [1, 60 \right ] M_{\odot}$, $m_2 \in \left [1, 3 \right ] M_{\odot}$, covering the region of mass-space thought to encompass BNSs and NSBHs; however, a mass-ratio cut $q \equiv m_1/m_2 \leq 20 $ is applied because reliable synthetic GW waveforms, which include higher modes, are currently unavailable for $q \gtrsim 20$. The binaries are distributed uniformly in co-moving volume {\it up to a limiting distance} $d_L^{\mathrm{max}}$, as well as in right-ascension ($\alpha \in \left [0, 2\pi \right]$), declination ($\sin \delta \in \left [-1, 1 \right ]$), polarization ($\psi \in \left [0, 2\pi \right]$), and inclination angle ($\cos \iota \in \left [-1, 1 \right]$). \footnote{Of the $100,000$ events, only those were selected that crossed an SNR threshold of 8. It turns out, however, since the distances are relatively small, that all events for the considered observing scenarios, cross this threshold. The only exception is the O5 scenario assuming a limiting distance of $200$ Mpc, where out of the $100,000$ events, only $\sim 30$ are undetected}. For each binary coalescence event considered, we compute the localisation skyarea for three fiducial times to coalescence comparable to typical slew-times of electromagnetic telescopes: $20,~40,~60$ seconds. We do so using $\ell=m=2$ mode waveforms, as well as those that include the $\ell=m=3$ and $\ell=m=4$ modes in addition to the $\ell=m=2$ mode, and evaluate the skyarea reduction factor due to the inclusion of the higher modes. We repeat this exercise for three limiting distances: $d_L^{\mathrm{max}} = 40,~100,~200$ Mpc \footnote{Note that we consider $100,000$ binaries for each of the limiting distances. This is to ensure that we are not statistically limited at lower distances.}, as well as the two observing scenarios: O5 and Voyager described in Section~\ref{sec:method}. 

The results are summarized via cumulative histograms plotted in Fig.~\ref{fig:cum-hist-all}. When higher modes are included, the localisation area was reduced by more than a factor of two for $\sim 30\% (40\%)$ of the binaries in O5 (Voyager) for early-warning times of $40-60$ sec, and a limiting distance of 40 Mpc. In the O5 scenario, up to $\sim 95\%$ ($80\%$), $\sim 60\%$ ($50\%$) and $\sim 35\%$ ($30\%$) events have localization skyareas less than $1000$ deg. sq. $20$s, $40$s, and $60$s before the merger respectively when higher modes are included (not included). These numbers increase to $\sim 100\%$ ($90\%$), $\sim 80\%$ ($60\%$) and $\sim 55\%$ ($40\%$) in the Voyager scenario. 

{\begin{figure*}
  \includegraphics[width=0.92\textwidth]{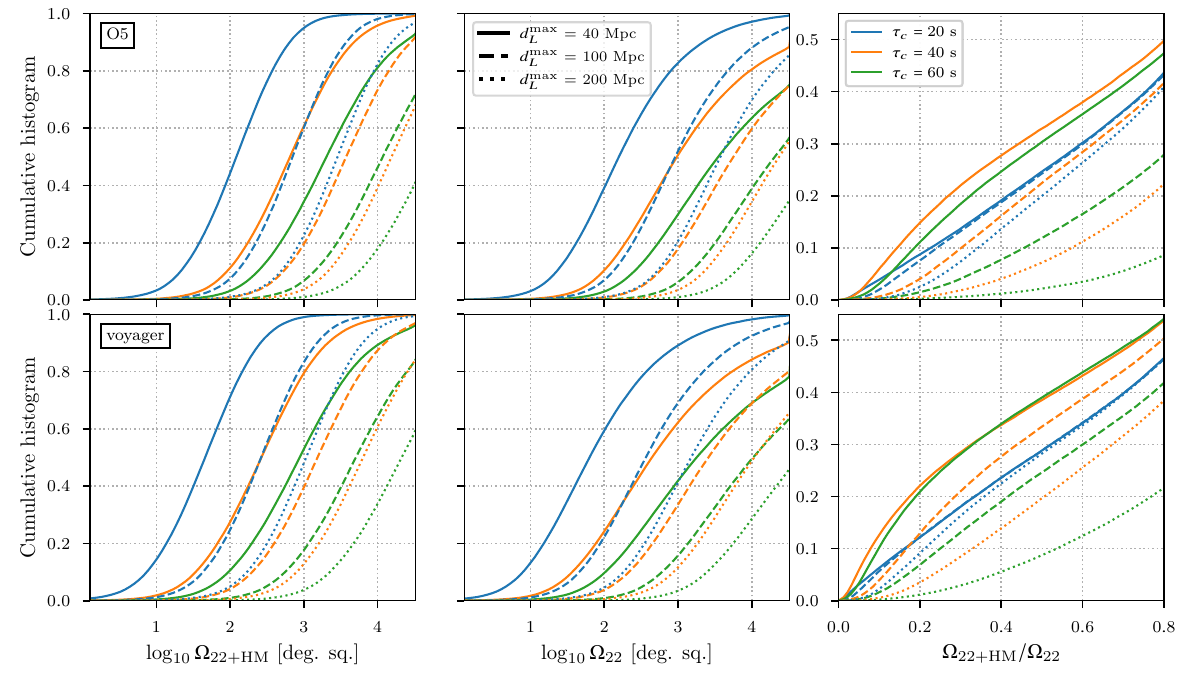}
  \caption{Cumulative histograms of localization skyareas for the simulated population of compact binaries $20$ sec. (blue), $40$ sec. (orange), and $60$ sec. (green) before the merger. {\itshape left:} localization skyarea using higher modes in addition to the dominant mode. {\itshape{middle:}} the same while using only the dominant mode. {\itshape{right:}} ratio of skyareas with/without higher modes. The solid, dashed and dotted lines correspond to three limiting distances $40$, $100$ and $200$ Mpc, respectively, while distributing the compact binaries uniformly in comoving volume. While employing higher modes for localization skyarea, up to $\sim 30\%$ and $\sim 40\%$ events have a reduction factor $\gtrsim 2$ (skyarea ratio $\leq 0.5$) in O5 and Voyager scenarios respectively for early-warning times of $40-60$ sec. In the O5 scenario, at best $\sim 95\%$ ($80\%$), $\sim 60\%$ ($50\%$) and $\sim 35\%$ ($30\%$) events have localization skyareas less than $1000$ deg. sq. $20$s, $40$s, and $60$s before the merger respectively when including (not including) higher modes. These numbers increase to $\sim 100\%$ ($90\%$), $\sim 80\%$ ($60\%$) and $\sim 55\%$ ($40\%$) in Voyager.}
    \label{fig:cum-hist-all}
\end{figure*}

\begin{figure*}
  \includegraphics[width=0.92\textwidth]{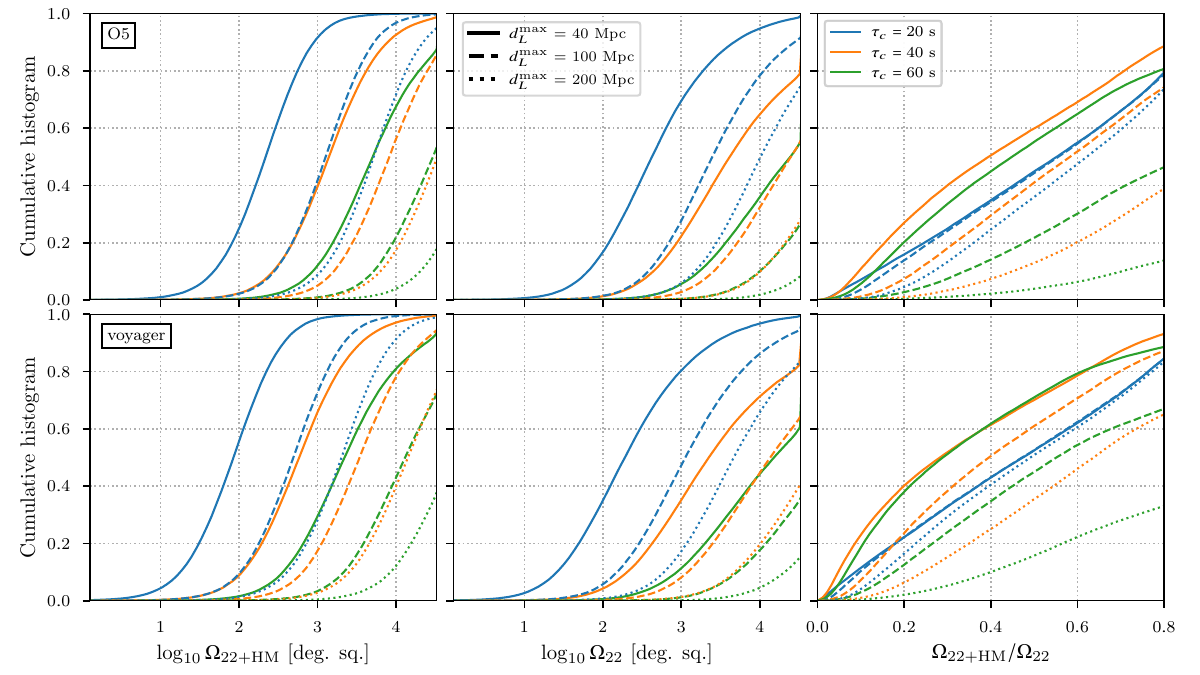}
  \caption{Same as Fig. \ref{fig:cum-hist-all}, but for the $\sim 55\%$ NSBH systems from the total population of 100,000 compact binaries. NSBH systems are selected from the total population such that $m_1 \geq 5.0 M_{\odot}$. Employing higher modes to determine localization skyareas results in up to $\sim 60\%$ and $\sim 70\%$ of the NSBH systems having skyarea reduction factors $\gtrsim 2$ for O5 and Voyager scenario respectively as compared to only dominant mode. In the O5 scenario, at best $\sim 90\%$ ($70\%$), $\sim 40\%$ ($20\%$) and $\sim 10\%$ ($ >10\%$) events have localization skyareas less than $1000$ deg. sq. $20$s, $40$s, and $60$s before the merger respectively when including (not including) higher modes. These numbers increase to $\sim 100\%$ ($80\%$), $\sim 65\%$ ($35\%$) and $\sim 30\%$ ($10\%$) in Voyager.}
    \label{fig:cum-hist-nsbh}
\end{figure*}}

\begin{figure*}
  \includegraphics[width=0.95\textwidth]{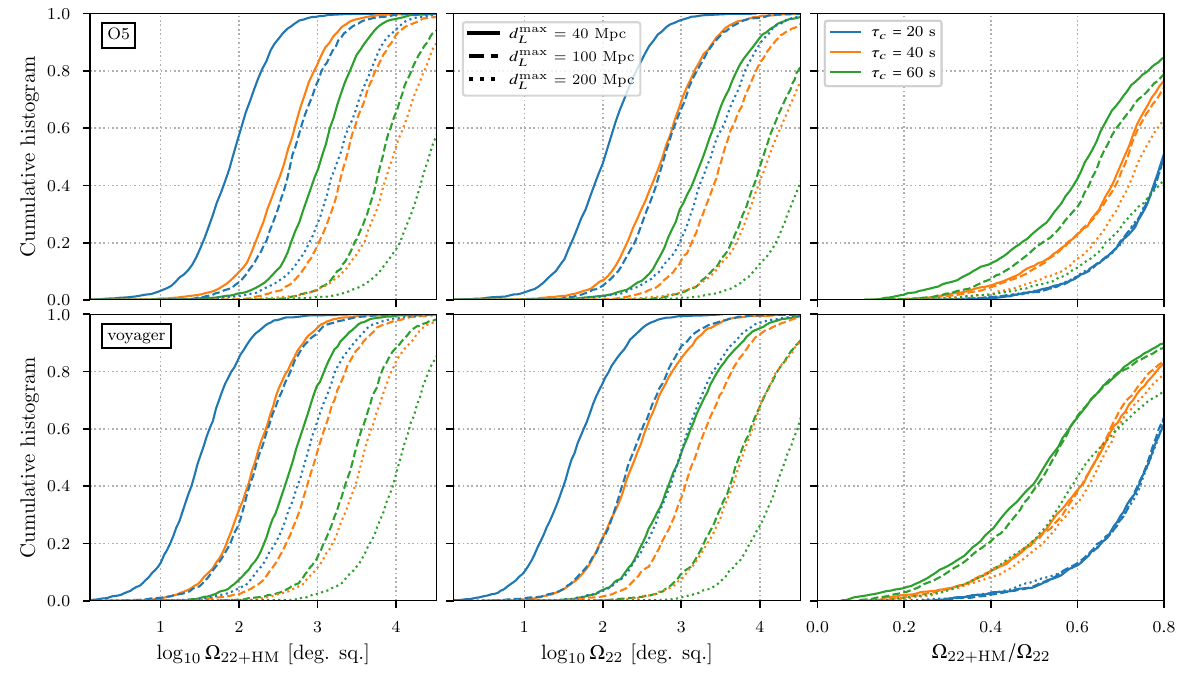}
    
  \caption{Same as in Fig. \ref{fig:cum-hist-nsbh}, except that here we consider only EM-Bright systems. The EOS is assumed to be AP4, and the spin distribution for the primary component is assumed to be isotropic. Out of total NSBHs ($\sim 55\%$), $\sim 2\%$ systems are found to be EM-Bright among which $\sim 20\%$ and $\sim 40\%$ events have localization skyarea improved by a factor $\gtrsim 2$ for O5 and Voyager  respectively while using higher modes. In the O5 scenario, at best $\sim 100\%$ ($>95\%$), $\sim 80\%$ ($70\%$) and $\sim 45\%$ ($30\%$) events have localization skyareas less than $1000$ deg. sq. $20$s, $40$s, and $60$s before the merger respectively while using (not using) higher modes. These numbers increase to $\sim 100\%$ ($100\%$), $\sim 90\%$ ($80\%$) and $\sim 75\%$ ($50\%$) in Voyager.}
   \label{fig:cum-hist-embrights-AP4}
\end{figure*}

\begin{figure*}
  \includegraphics[width=0.95\textwidth]{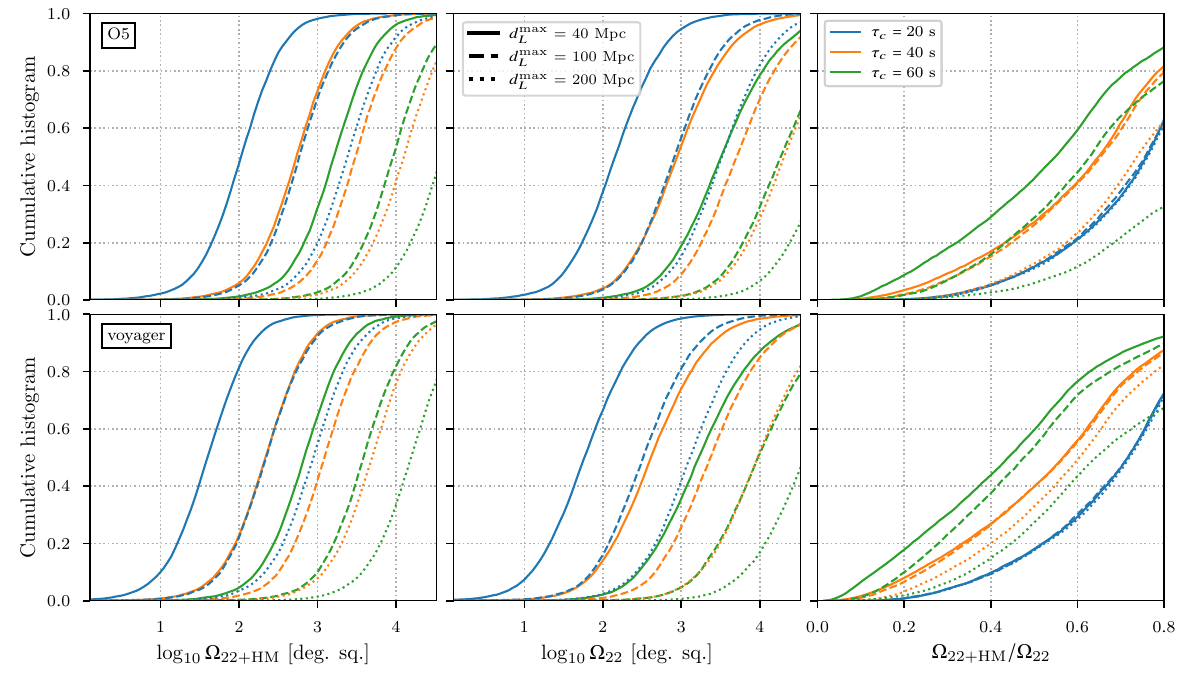}
  \caption{Same as Fig. \ref{fig:cum-hist-embrights-AP4} with the isotropic distribution of the primary's spin changed to an aligned one. In this case, out of the total NSBHs, $\sim 15\%$ systems are found to be EM-Bright among which $\sim 40\%$ and $\sim 55\%$ events have localization skyarea improvement factor $\gtrsim 2$ for O5 and Voyager respectively while using higher modes. In the O5 scenario, at best $\sim 97\%$ ($95\%$), $\sim 75\%$ ($55\%$) and $\sim 30\%$ ($20\%$) events have localization skyareas less than $1000$ deg. sq. $20$s, $40$s, and $60$s before the merger respectively while using (not using) higher modes. These numbers increase to $\sim 100\%$ ($>97\%$), $\sim 90\%$ ($75\%$) and $\sim 65\%$ ($40\%$) in Voyager.}
   \label{fig:cum-hist-embrights-AP4-aligned-spin}
\end{figure*}

\begin{figure*}
  \centering
  \subfloat[]{
    \includegraphics[width=0.95\textwidth]{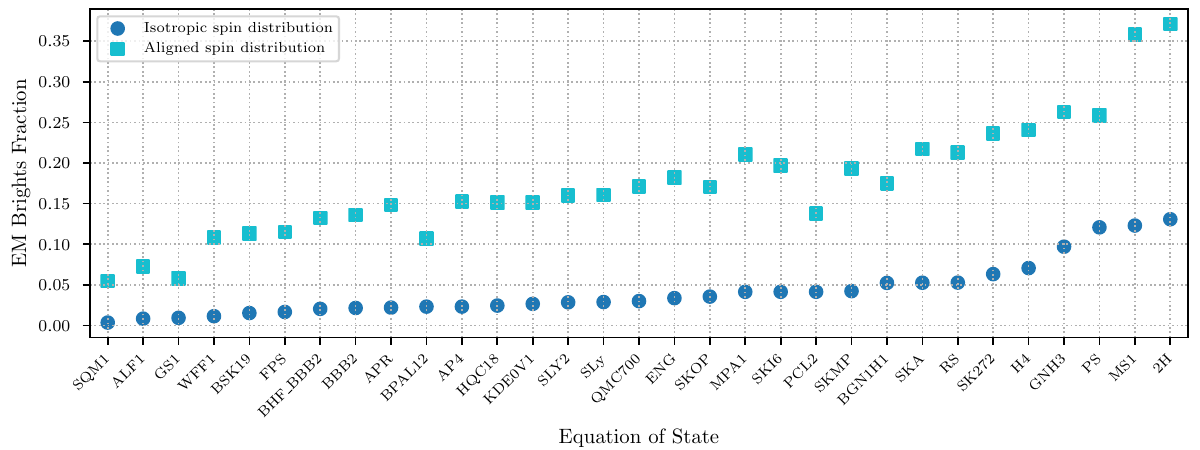}
    \label{fig:fraction-eos}}\\
  \subfloat[]{
    \includegraphics[width=0.95\textwidth]{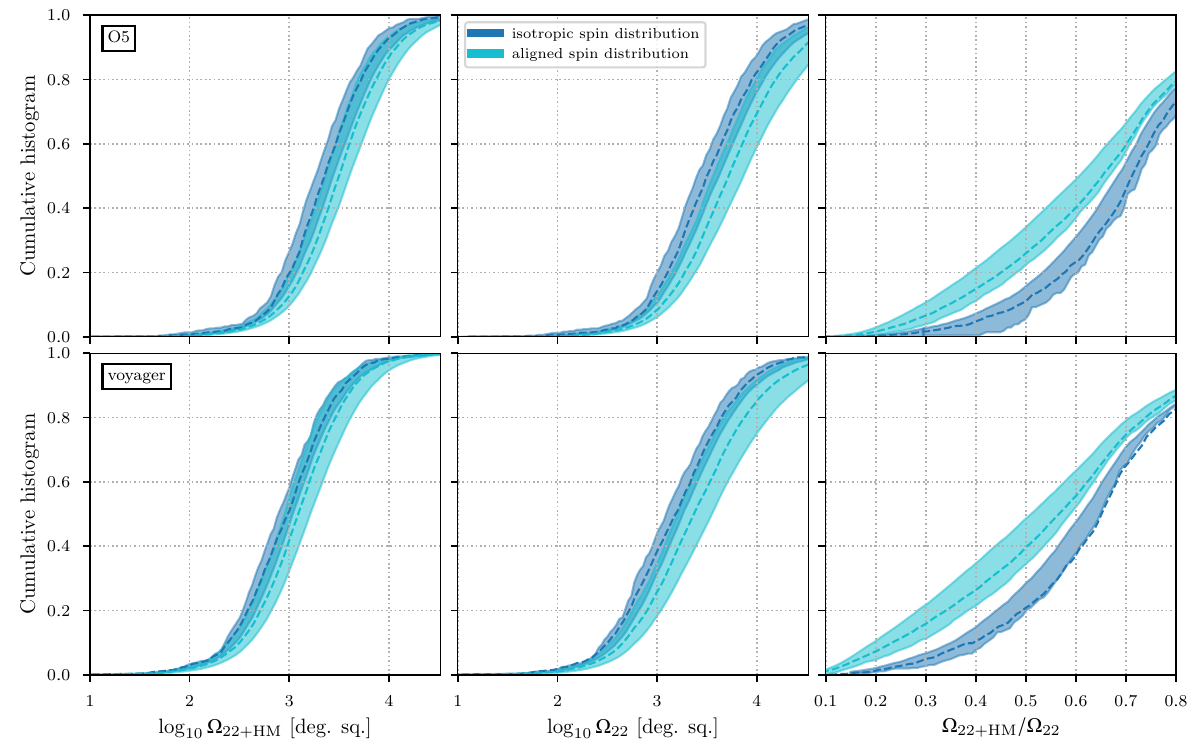}
    \label{fig:cum-hist-embrights-all-EOS}}
  \caption{\ref{fig:fraction-eos}: Fraction of the NSBH population that are EM-Bright, for isotropic and aligned BH-spin distributions and a range of EOSs \citep{GW170817-SOURCE-PROPERTIES}. These fractions vary from $\sim 0.5\% - 10\%$ for the isotropic spin distribution, and $\sim 5\% - 35\%$ for the aligned spin distribution. \ref{fig:cum-hist-embrights-all-EOS}: Same as Fig.~\ref{fig:cum-hist-embrights-AP4}, but for a range of EOSs (identical to those in Fig.~\ref{fig:fraction-eos}), isotropic and aligned BH-spin distributions, and only one limiting distance ($d_L^{\mathrm{max}} = 100$ Mpc) and early-warning time ($40$ sec.). The shaded band covers all the EOSs considered, and the dashed lines correspond to the AP4 EOS (cf. Figs.~\ref{fig:cum-hist-embrights-AP4} and \ref{fig:cum-hist-embrights-AP4-aligned-spin}). The fraction of the EM-Bright systems for which the localisation skyarea reduction factor $\gtrsim 2$ is greater by $20\%$ for the aligned spin distribution than for the isotropic spin distribution. On the other hand, the variation of this fraction due to different EOSs is only $\sim 10\%$.}
\end{figure*}

Of the 100,000 binaries considered, $\sim 55\%$ are NSBHs \footnote{We set a conservative upper limit on the mass of the neutron star to be $3 M_{\odot}$, and a lower limit on the mass of the black hole to be $5 M_{\odot}$. The lower limit on the neutron mass is set to $1 M_{\odot}$.}. However, this fraction may change depending on the choice of astrophysical distribution of masses of binary systems. Since a larger fraction of these events have heavier masses, this population on average spends a shorter time in the frequency band of the detectors. As a result, the fraction of events with smaller localisation areas is reduced, as shown in Fig.~\ref{fig:cum-hist-nsbh}. On the other hand, a larger proportion of these events have asymmetric masses. Therefore the fraction of these events for which the reduction factor is $ \gtrsim 2$ increases significantly. For example, for the O5 (Voyager) scenario, assuming an early-warning time of $40$ seconds and a limiting distance of $40$ Mpc, this fraction is $\sim 60\% \left (70 \% \right)$. In fact, we see that for $30\%$ ($40\%$) of the events, the sky area is improved by more than a factor of $4$. The fraction of events with high reduction factors decrease for other early warning times, as well as larger limiting distances. Nevertheless, this fraction is significantly higher in general than for the total population, which also includes nearly-symmetric-mass BNS systems. This highlights the power of using higher modes for early-warning of asymmetric mass systems.
 
We then focus on potentially EM-Bright binaries among the NSBH systems. Determining whether an NSBH system will produce an EM counterpart is in general complicated and arguably still an open question. However, the expectation is that a system that produces a counterpart will also produce a post-merger remnant baryonic mass \citep{FoucartEMB1, FoucartEMB2} \footnote{Note that there are models of NSs \citep{InspiralCounterpart} that predict the emission of a counterpart before the NSBH (or BNS) merger, and therefore before a remnant baryonic mass gets produced. We use the \textsc{p-astro} PyPI sofware package \citep{p-astro} to determine whether an NSBH is EM-Bright.}, which we take to be a proxy for the existence of the counterpart.

While the early warning times and localisation skyareas early in the inspiral have negligible dependence on the component spins of the binary, the production of the remnant baryonic mass (and thus potential EM-Brightness) is crucially dependent on the spin of the black hole, as well as the EOS of the neutron star. We consider a range of EOSs that include both stiff and realistic (consistent with GW170817 \citep{GW170817-SOURCE-PROPERTIES}) ones.  We also choose two spin distributions: an isotropic distribution (spin magnitudes uniformly distributed between 0 and 1 and spin angles isotropically distributed) and an aligned distribution (same distribution of spin magnitudes, but spin vector always aligned with the orbital angular momentum of the binary).

Of the total NSBH binaries ($55\%$ of the total population), $2\%$ ($15\%$) are EM-Bright for the AP4 EOS \citep{AP4} assuming an isotropic (aligned) spin distributions \footnote{These fractions might change depending on the choice of astrophysical mass distribution.} The isotropic-spin (aligned-spin) population producing small (large) fraction of EM-Bright binaries is consistent with our expectation --- smaller (larger) aligned spin black holes have bigger (shorter) radii for their innermost stable circular orbits (ISCO), and hence will result in a smaller (larger) fraction of systems with unbound tidal/merger ejecta. The improvements in the sky localisation for these EM-Bright populations are plotted in Figs.~\ref{fig:cum-hist-embrights-AP4} and \ref{fig:cum-hist-embrights-AP4-aligned-spin}. The fraction of these EM-Bright systems that produce skyarea reduction factors $\gtrsim 2$ are reduced with respect to the same fraction for the total NSBH population, for the isotropic spin distribution. This is not entirely unexpected -- unbound tidal/merger ejecta is produced predominantly for nearly equal-mass binaries \citep{FoucartEMB1}, where the improvements due to higher modes are relatively modest.  Nevertheless, for the Voyager scenario, one finds that the fraction of events with reduction factor $\gtrsim 2$ can be as large as $\sim 40\%$. For the aligned spin distribution, this fraction can be as large as $ \sim 40\% (60\%)$ in O5 (Voyager). These estimates pertain to the case when all binary systems are located within a limiting distance of 40 Mpc. Comparable results are also found for a limiting distance of $100$ Mpc.

In Fig.~\ref{fig:fraction-eos} and \ref{fig:cum-hist-embrights-all-EOS}, we study the effect of varying the EOS on the fraction of the NSBH population distributed upto a limiting distance of $100$ Mpc that are EM-Bright, and on the fraction of EM-Bright systems that provides significant early-warning gains upon inclusion of the higher modes. We consider $31$ EOSs (as implemented in the \textsc{LALSuite} \citep{lalsuite} software package), and find that the EM-Bright fraction varies from $\sim 0.5-14\%$ for isotropic spins, and $\sim 5 - 35\%$ for aligned spins. On the other hand, varying the EOS changes the fraction of EM-Bright systems that produce reduction factors $\gtrsim 2$ by $\sim 10 \%$.

\begin{figure}
  \begin{center}
    \includegraphics[width=1\columnwidth]{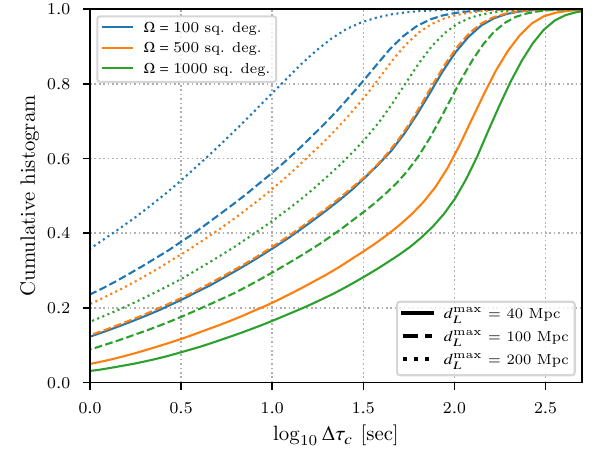} 
  \end{center}
  \caption{ The early-warning time gains are shown while localizing compact binaries for three different skyareas of $100$ (blue), $500$ (orange) and $1000$ (green) deg. sq. in 3G scenario. We get time gains upto several minutes while employing higher modes in localization as compared to when we would have used only quadrupole mode. Again, the analysis has been repeated for three upper limits on distance.
  }
  \label{fig:cum-hist-delta-tc-3G}
\end{figure}

We also evaluate the use of higher modes for early-warning for the 3G network of detectors described in Sec.~\ref{sec:method}. We compute the early-warning time gained with the inclusion of higher modes, for three values of the localisation skyarea: $100$, $500$ and $1000$ sq. deg, and the full population of 100,000 binaries that include both BNS and NSBH systems. The results are summarized in Fig.~\ref{fig:cum-hist-delta-tc-3G}. For a limiting distance of $40$ Mpc, $\sim 60 \%$ of the events have early-warning time gains of over a minute for a $1000$ sq. deg localization area and above $40$ sec. for a localization area of $500$ sq. deg. Even at 100 Mpc, the median improvement is about $30$ seconds for a $500$ sq. deg. localization area.

\begin{figure*}
  \centering
  \subfloat[]{    \includegraphics[width=0.95\textwidth]{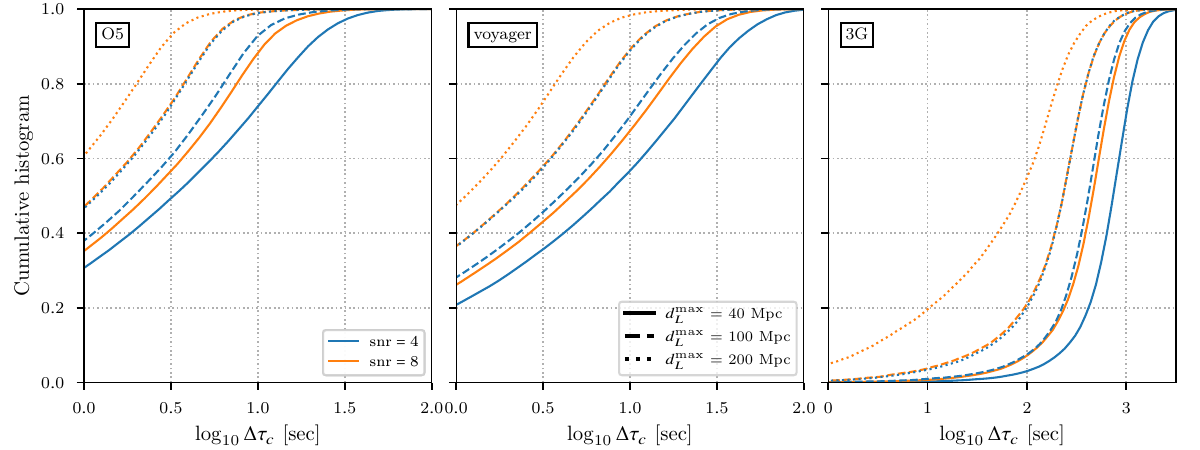}
    \label{fig:cum-hist-delta-tc-all}}\\
  \subfloat[]{ \includegraphics[width=0.95\textwidth]{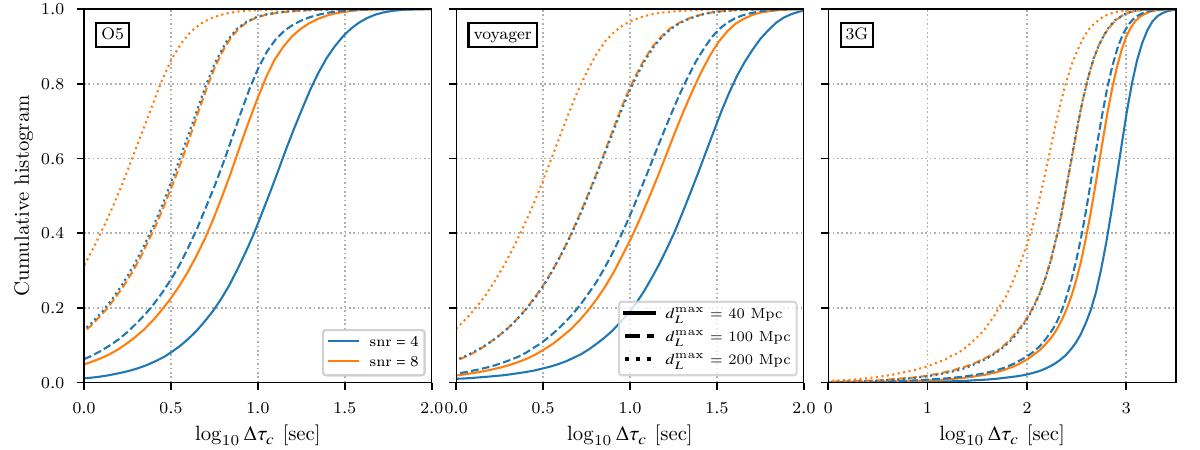}
    \label{fig:cum-hist-delta-tc-nsbh}}
  \caption{\ref{fig:cum-hist-delta-tc-all}: Gains in early warning times, given fiducial SNRs, for the full population, as well as three scenarios and limiting luminosity distances. Up to $\sim 30\%$ of the binaries have gains of greater than $10$ sec. in O5. This increases to $\sim 40 \%$ in Voyager. In the 3G scenario, up to $80\%$ of the events have time gains $> 500$ sec.
    \ref{fig:cum-hist-delta-tc-nsbh}: Gains in early warning times, given fiducial SNRs, for the NSBH population, as well as three scenarios and luminosity distances. Up to $\sim 60\%$ of the binaries have gains of greater than $10$ sec. in O5. This increases to $\sim 80 \%$ in Voyager. In the 3G scenario, up to $80\%$ of the events have time gains $> 500$ sec. In the 3G scenario, the gains are similar to that of the full population, except for $d_L^{\mathrm{max}}= 200$ Mpc. where there are $\sim 10\%$ more events with time gains $> 100$ sec.
  }
\end{figure*}

Even if an early-warning skyarea is not sufficiently small to allow for small-field-of-view telescopes to rapidly cover, detectors that only need a broad localisation area would benefit from an early-warning detection of the event well before the merger. To that end, we compute the gain in early-warning times with the inclusion of higher modes, for fiducial values of the network SNR (SNR = 4, 8). In Fig.~\ref{fig:cum-hist-delta-tc-all}, we consider the full population of 100,000 binaries. Up to $\sim 30\%$ of the binaries have gains of greater than $10$ sec. in O5. This increases to $\sim 40 \%$ in Voyager. In the 3G scenario, up to $80\%$ of the events have time gains $> 500$ sec. In Fig.~\ref{fig:cum-hist-delta-tc-nsbh}, we consider the population of $\sim 55\%$ NSBH systems. Up to $\sim 60\%$ of the NSBHs have gains of greater than $10$ sec. in O5. This increases to $\sim 80 \%$ in Voyager. In the 3G scenario, up to $80\%$ of the events have time gains $> 500$ sec. In the 3G scenario, the gains are similar to that of the full population, except for $d^{\mathrm{max}}_L= 200$ Mpc. where there are $\sim 10\%$ more events with time gains $> 100$ sec.

\section{Summary and Outlook}\label{sec:conclusion}
	BNS and NSBH mergers, which have the potential to produce EM-counterparts, have at best been confidently detected only a handful of times. Models of such systems, especially related to the EM emission, therefore remain unconstrained by observational data. The relative rarity of such events, compared to BBH mergers (which were an almost regular occurrence in O3), as well as the complex physics involved, motivates the need for GW early-warning of such events. 

The LIGO/Virgo collaboration has already commissioned a trial system that is able to relay the discovery of BNSs up to tens of seconds pre-merger on making use of the $\sim 100$ sec. time evolution of such low-mass binaries \citep{user_guide_ew}. Current large field of view instruments like the Zwicky Transient Facility (ZTF) (~47 sq. degs.) have the potential to follow-up $\mathcal{O}$(100 sq. degs.) sky localizations in a few pointings. Given the few degrees per second slew times \citep{ZTF}, early warning triggers present a unique opportunity for such instruments to begin scheduling operations from GW events before the merger. Next generation instruments like the Vera Rubin Observatory (VRO) \citep{RubinObs} will reach unprecedented depths in the follow-up of EM-GW counterparts. Early warning will open an entirely new area to early photometric and spectroscopic observations of the prompt high energy physics in the EM spectrum \citep{Metzger2019}.

While research towards GW early-warning for BNS systems is currently being pursued, early-warning for NSBH systems remains relatively un-explored. Our previous work \citep{Kapadia2020} demonstrated the reduction in localisation skyareas upon the inclusion of higher order modes in addition to the dominant mode currently used in templated real-time GW searches. This reduction is significant for a range of asymmetric mass systems and inclination angles, a fraction of which could produce EM-counterparts under different assumptions of the EOS, and moderately spinning BH component. 

In this work, we determine, in a statistical sense, the early-warning gains upon inclusion of higher modes, for a fiducial population of 100,000 binaries that include BNS and NSBH systems. We consider component masses, log-uniformly distributed, spanning $m_1 \in \left [1, 60 \right ] M_{\odot}$ and $m_2 \in \left [1, 3 \right] M_{\odot}$ with an applied mass ratio cut of $q \leq 20$. The source location is isotropically distributed, as well as uniformly in co-moving volume. A significant fraction of events have improved localisation by a factor $\gtrsim 2$. This fraction can be as high as $\sim 30 \%$ and $\sim 40 \%$ for a limiting distance of $40$ Mpc in O5 and Voyager respectively (Fig.~\ref{fig:cum-hist-all}).

These fractions improve by over $\sim 10 \%-40\%$ when only NSBHs are considered, but remain relatively unchanged when only EM-Bright sub-populations of the NSBH population are considered, assuming an isotropic spin distribution. However, for aligned spin distributions, the fractions are only marginally worse than those for the total NSBH population. By varying the EOS to encompass a range of siffnesses, we find that the fraction of NSBH systems that are EM-Bright can vary by up to $30\%$ for the aligned spin distribution, and the fraction of EM-Bright events with reduction factor $\gtrsim 2$ varies by $\sim 10\%$ for both the aligned and isotropic spin distributions.

Translating fraction of events where early-warning gains due to higher modes are significant, to an actual frequency during an observing run, is straightforward provided the true rate of these events is known. Unfortunately, given the paucity of BNS and NSBH detections, the rates of these events to date remain highly uncertain. Nevertheless, as an example, let us consider the NSBH rate upper limit of $\sim600\ \mathrm{Gpc}^{-1}\textrm{yr}^{-1}$ \citep{GWTC-1}, which corresponds to about $5$ detections per year within $200$ Mpc, assuming all NSBH mergers within the corresponding volume are detected, as would be expected in the Voyager scenario. From Fig.~\ref{fig:cum-hist-nsbh}, $\sim 30\%$ of the NSBH population have a reduction factor $\gtrsim 2$ for an early-warning time of $40$ sec, which corresponds to $\sim 3$ NSBH detections in 2 years. 

While higher modes may provide significant reduction factors, these may not always result in tight skyarea localisations. Nevertheless, even though skyareas of several hundreds of square degrees may not be covered by a single telescope sufficiently rapidly to capture the transient event, a joint effort involving multiple telescopes could do so, especially if assisted by a galaxy catalog. Employing a hierarchical tiling strategy that prioritizes regions of the skymap with higher probability values as well as galaxy locations within the skymap, multiple telescopes could cover a large skyarea efficiently enough to capture the EM-counterpart at its onset \citep{SlewingStrategy}. One might also envision an optimal set-up involving a coordianted effort between the GW network of detectors, and a global network of EM-telescopes (see, for example, \cite{Grandma}), where the evolving and shrinking GW skymaps are continously streamed to robotic/automated telescopes that continously track them. 

Sky maps that are $\mathcal{O}$(1000) sq. deg. wide could also be exploited by large field of view telescopes. The Swift gamma-ray burst (GRB) satellite has a large field of view component, BAT (Burst Alert Telescope), which covers $\sim 4600$ sq. deg. A GRB detected by BAT triggers the XRT (X-Ray Telescope) and UVOT (Ultra-Violet and Optical Telescope) components with significantly smaller fields of view, and slew-times of 20-75 sec, for follow-up. \citep{Swift}. A skymap with an area a factor of a few smaller than the BAT field of view could allow XRT and UVOT to start slewing towards the location of the counterpart before its onset, potentially gaining tens of seconds to a minute of early-warning time. In fact, if BAT is itself pointing in a direction that mostly excludes the early-warning GW-skymap, it could start slewing towards the localisation region so as to encompass the maximum probability region within its field of view before the onset of the counterpart. Similar strategies can also be applied for the Astrosat telescope \citep{Astrosat}.

The Konus-Wind (KW) gamma-ray burst spectrometer \citep{Konus-Wind} could take advantage of an early-warning detection, even though its omnidirectional instruments are unlikely to benefit from an early-warning skymap. KW records triggers above some predefined threshold. Thus, an algorithm that triggers on GW early-warning detection could ensure that subthreshold GRB triggers over the duration of the CBC transient could be recorded for subsequent offline analyses. 

Radio telescopes are also likely to benefit from early-warning of GW events. The MWA radio telescope has a 600 sq. deg. field of view and can start following up events $\sim 20$ sec. after its detection \citep{Kaplan2015}. Therefore even a $\mathcal{O}$(1000) sq. deg. skymap at an early-warning time of $60$ sec. would save tens of seconds, since MWA should be able to cover this area in a few pointings. The ASKAP radio telescope has a significantly smaller field of view ($\sim 30$ sq. deg.), and a slewing rate of a few degrees per second, and could therefore take advantage of a $\mathcal{O}$(100) sq. deg. skymap with tens of seconds of early-warning time \citep{Dobie2019}. Telescopes with much larger fields of view $\mathcal{O}$(10,000) sq. deg. buffer data periodically. A triggering algorithm, based on the GW early-warning, could ensure that the relevant data is recorded for offline analyses.

While localisation skymap is arguably the most important information for telescopes to follow-up a GW early-warning event, other information such as estimates of the luminosity distance, inclination angle, mass-ratio and spin could also allow astronomers to determine whether the event is worthy of follow-up. For example, the optical telescopes have a limiting distance to which they can view an event with short exposure times; furthermore, better estimates of the mass-ratio and spin could better help determine if the event is EM-Bright. Inclusion of higher modes will likely improve the estimation of these parameters, although a detailed investigation for a realistic population needs to be conducted. We leave this for future work.

\section*{Acknowledgements}
We thank Surabhi Sachdev and Tito Dal Canton for carefully reviewing our manuscript and providing useful comments. We also thank Shaon Ghosh and Leo P. Singer for illuminating discussions. We are grateful to Stephen Fairhurst for clarifications on \citep{Fairhurst1, Fairhurst2}, Srashti Goyal for help with the \textsc{PyCBC} \citep{pycbc_software} implementation of the antenna pattern functions, and Varun Bhalerao and Shabnam Iyyani for suggestions on possible uses of early warning for gamma-ray and radio telescopes. SJK's, MKS's, MAS's and PA's research was supported by the Department of Atomic Energy, Government of India. In addition, SJK's research was funded by the Simons Foundation through a Targeted Grant to the International Centre for Theoretical Sciences, Tata Institute of Fundamental Research (ICTS-TIFR). PA's research was funded by the Max Planck Society through a Max Planck Partner Group at ICTS-TIFR and by the Canadian Institute for Advanced Research through the CIFAR Azrieli Global Scholars program. DC is supported by the Illinois Survey Science Fellowship of the Center for AstroPhysical Surveys at the University of Illinois at Urbana-Champaign. DC would like to thank the ICTS-TIFR for their generous hospitality; a part of this work was done during his visit to ICTS. Numerical calculations reported in this paper were performed using the Alice cluster at ICTS-TIFR.
\section*{Data Availability}
The data underlying this article will be shared on reasonable request to the corresponding author.


\end{document}